\documentclass[10pt,a4paper,twoside]{article}

\usepackage{indentfirst}                        
\usepackage{graphicx}                           
\usepackage{amsgen,amsfonts,amssymb,amsbsy}     

\newenvironment{resum}{\begin{quote}\small}{\end{quote}}

\newcommand{\bfsf}[1]{\textsf{\textbf{#1}}}

\begin{document}

\thispagestyle{plain}           

\begin{center}


{\LARGE\bfsf{Non-linear relativistic perturbation theory \\
with two parameters}}

\bigskip


\textbf{Carlos F. Sopuerta}$^1$, \textbf{Marco Bruni}$^1$ and \textbf{Leonardo
Gualtieri}$^2$


$^1$\textsl{University of Portsmouth, United Kingdom.} \\
$^2$\textsl{Universit\`a di Roma ''La Sapienza'' and Sezione INFN
  ROMA, Italy.}

\end{center}

\medskip


\begin{resum}
  An underlying fundamental assumption in relativistic perturbation
  theory is the existence of a parametric family of spacetimes that
  can be Taylor expanded around a background.  Since the choice of the
  latter is crucial, sometimes it is convenient to have a perturbative
  formalism based on two (or more) parameters.  A good example is the
  study of rotating stars, where generic perturbations are constructed
  on top of an axisymmetric configuration built by using the slow
  rotation approximation.  Here, we discuss the gauge dependence of
  non-linear perturbations depending on two parameters and how to
  derive explicit higher order gauge transformation rules.
\end{resum}

\bigskip



\section{Introduction}

Relativistic perturbation theory is based on the assumption that
there exists a parametric family of spacetimes that can be
Taylor expanded around a certain background.  The perturbations are then
defined as the derivative terms of this series, evaluated on that
background. In most cases of interest one deals with an expansion
in a single parameter, which can either be a formal one, as in
cosmology~\cite{BMMS}, or can have a specific physical meaning,
as in the study of binary black hole mergers via the close limit
approximation~\cite{closelimit}.  In some physical applications
it may be instead convenient to construct a perturbative formalism
based on two (or more) parameters, because the choice of background
is crucial in having a manageable theory.  The study of perturbations
of stationary axisymmetric rotating stars is a good example.
In this case, an analytic stationary axisymmetric solution
is not known, at least for reasonably interesting equations of state.
A common procedure is to treat axisymmetric stars using the
so--called slow rotation approximation, so that the background is a
star with spherical symmetry. In this approach
the first order in $\Omega$ discloses frame dragging effects, with the
star actually remaining spherical; $\Omega^2$ terms carry the effects of
rotation on the fluid.  Generic time dependent perturbations of the
rotating star (parametrized by a dummy parameter $\lambda$ and
describing oscillations) are then built on top of the stationary axisymmetric
perturbations in $\Omega$.  Clearly, in this approach any interesting physics
requires non--linear perturbations, as at least terms of order
$\lambda\Omega$ need to be considered.  Here, we discuss the
formulation of non-linear relativistic perturbation theory with two parameters.
Details can be found in~\cite{BGS}.


\section{Two-parameter perturbation theory}

We start with the introduction of an adequate framework for the 
formulation of perturbation theory with two parameters.
Following the spirit of~\cite{stewart} we introduce an $(m+2)$--dimensional
manifold ${\cal N}$, foliated by $m$--dimensional submanifolds diffeomorphic to
${\cal M}$, the spacetime manifold, so that ${\cal N}={\cal M}\times{\rm 
I\!R}^2$.
We shall label each copy of ${\cal M}$ by the corresponding value of the 
parameters
$\lambda,\Omega$.  The manifold ${\cal N}$ has a natural differentiable
structure which is the direct product of those of ${\cal M}$ and
${\rm I\!R}^2$.  We can then choose charts on ${\cal N}$ in which $x^{\mu}$
($\mu=0,1,\dots m-1$) are coordinates of each leaf
${\cal M}_{\lambda,\Omega}$ and $x^m=\lambda,\,x^{m+1}=\Omega$.

If a tensor field $T_{\lambda,\Omega}$ is given
on each ${\cal M}_{\lambda,\Omega}$, we have that a tensor field $T$ is
automatically defined on ${\cal N}$ by the relation
$T(p,\lambda,\Omega):= T_{\lambda,\Omega}(p)$, with
$p\in{\cal M}_{\lambda,\Omega}$.
In particular, on each ${\cal M}_{\lambda,\Omega}$ one has a
metric $g_{\lambda,\Omega}$ and a set of matter fields
$\tau_{\lambda,\Omega}$, satisfying a set of field equations
\begin{eqnarray} 
{\cal E}[g_{\lambda,\Omega},\tau_{\lambda,\Omega}]=0\,, \nonumber
\end{eqnarray}
which we do not need to specify since this formulation of perturbation theory 
can be applied to any spacetime theory.

We now want to define the perturbation in any tensor field $T$,
therefore we must find a way to compare
$T_{\lambda,\Omega}$ with the background value $T_0$.  This requires
a prescription for identifying points of ${\cal M}_{\lambda,\Omega}$
with those of ${\cal M}_0$. This is easily accomplished by assigning a
diffeomorphism
$\varphi_{\lambda,\Omega}:{\cal N}\rightarrow{\cal N}$ such that
$\left.\varphi_{\lambda,\Omega}\right|_{{\cal M}_0}:{\cal M}_0
\rightarrow{\cal M}_{\lambda,\Omega}$.
Clearly, $\varphi_{\lambda,\Omega}$ can be regarded as the
member of a two--parameter group of diffeomorphisms $\varphi$ on ${\cal N}$,
corresponding to the values of $\lambda,\Omega$ of the group
parameter. Therefore, we could equally well give the
vector fields $^{\varphi}\eta,\,^{\varphi}\zeta$ that generate
$\varphi$.  In our construction the group of diffeomorphisms is choosen
to be Abelian and therefore $^{\varphi}\eta$ and $^{\varphi}\zeta$
commute.   In this way, perturbing first with respect to the parameter
$\lambda$ and afterwards with respect to $\Omega$ coincides with the
converse operation.

The choice of the group of diffeomorphisms $\varphi$, or
equivalently, the choice of the vector fields 
$^{\varphi}\eta,\,^{\varphi}\zeta$,
is what in perturbation theory is called a {\em gauge choice}.
Using the pull-back of $\varphi$, the perturbation in $T$ can now be 
defined simply as
\begin{eqnarray}
\Delta_0
T_{\lambda,\Omega} := \left.\varphi^*_{\lambda,\Omega}T\right|_{{\cal M}_0}
-T_0\,.   \label{defperturbation}
\end{eqnarray}
The first term on the right--hand side can be Taylor--expanded to get~\cite{BGS}
\begin{eqnarray}
\Delta_0 T_{\lambda,\Omega}=\sum_{k,k'=0}^{\infty}
\frac{\lambda^k{\Omega^{k}}'}{k!\,k'!}\delta^{(k,k')}_\varphi T-T_0\,, 
\label{perturbationexpansion}
\end{eqnarray}
where
\begin{eqnarray}
\delta^{(k,k')}_\varphi T:=\left[\frac{\partial^{k+k'}}
{\partial \lambda^k\partial\Omega^{k'}}\varphi^*_{\lambda,\Omega}T
\right]_{\lambda=0,\Omega=0,{\cal M}_0}\!\!\! = \left.{\cal 
L}^k_{\,^{\varphi}\!\eta}
{\cal L}^{k'}_{\,^{\varphi}\!\zeta}T\right|_{{\cal M}_0}\,, 
\label{defperturbationk}
\end{eqnarray}
and here ${\cal L}$ denotes Lie differentiation.  This last expression
defines the perturbation of order $(k,k')$ of
$T$ (notice that $\delta^{(0,0)}_\varphi T=T_0$).  It is worth noticing
$\Delta_0 T_{\lambda,\Omega}$ and $\delta^{(k,k')}_\varphi T$ are
defined on ${\cal M}_0$; this formalizes the statement one
commonly finds in the literature that {\it perturbations are
fields living in the background}. It is important to
appreciate that the parameters $\lambda,\,\Omega$
labelling the various spacetime models also serve to
perform the expansion (\ref{perturbationexpansion}), and
therefore determine what one means by {\em perturbations of
order $(k,k')$}.


\section{Gauge invariance and gauge transformations}

Let us now suppose that two gauges $\varphi$ and $\psi$,
described by pairs of vector fields $(^{\varphi}\eta,\,^{\varphi}\zeta)$
and $(^{\psi}\eta,\,^{\psi}\zeta)$ respectively, are
defined on ${\cal N}$.
Correspondingly, the integral curves of
$(^{\varphi}\eta,\,^{\varphi}\zeta)$ and $(^{\psi}\eta,\,^{\psi}\zeta)$
define two two--parameter groups of diffeomorphisms $\varphi$ and
$\psi$ on ${\cal N}$, that connect any two leaves of the
foliation. Thus, $(^{\varphi}\eta,\,^{\varphi}\zeta)$
and $(^{\psi}\eta,\,^{\psi}\zeta)$ are everywhere transverse to
${\cal M}_{\lambda,\Omega}$ and points lying on the same
integral surface of either of the two are to be regarded
{\it as the same point} within the respective gauge:
$\varphi$ and $\psi$ are both point identification maps,
i.e.\ two different gauge choices.

The pairs of vector fields $(^{\varphi}\eta,\,^{\varphi}\zeta)$
and $(^{\psi}\eta,\,^{\psi}\zeta)$ can both be used to pull back a
generic tensor $T$ and therefore to construct two other
tensor fields $\varphi^*_{\lambda,\Omega}T$ and
$\psi^*_{\lambda,\Omega}T$, for any given value of
$(\lambda,\Omega)$. In particular, on ${\cal M}_0$ we now have
three tensor fields, i.e.\ $T_0$ and
\begin{eqnarray}
T^\varphi_{\lambda,\Omega}:= \left.\varphi^*_{\lambda,\Omega}
T\right|_{{\cal M}_0}\,,~~~~~~~~
T^\psi_{\lambda,\Omega}:= \left.\psi^*_{\lambda,\Omega}
T\right|_{{\cal M}_0}\,. \nonumber
\end{eqnarray}
Since $\varphi$ and $\psi$ represent gauge choices for mapping a perturbed
manifold ${\cal M}_{\lambda,\Omega}$ into the unperturbed one ${\cal M}_0$,
$T^\varphi_{\lambda,\Omega}$ and
$T^\psi_{\lambda,\Omega}$ are the representations, in ${\cal M}_0$, of the 
perturbed
tensor according to the two gauges. 
Using~(\ref{defperturbation}--\ref{defperturbationk})
we can write
\begin{eqnarray}
 T^\varphi_{\lambda,\Omega} & = &
\sum_{k=0}^{\infty}\frac{\lambda^k{\Omega^k}'}{k!\,k'!}
\delta^{(k,k')}_\varphi T=\sum_{k,k'=0}^{\infty}
\frac{\lambda^k{\Omega^k}'}{k!\,k'!}
{\cal L}^k_{\,^{\varphi\!}\eta}{\cal L}^{k'}_{\,^{\varphi\!}\zeta}T=T_0+
\Delta_0^{\varphi}T_{\lambda,\Omega}\,,\label{defexpX}
\end{eqnarray}
and the analogous one  for $T^\psi_{\lambda,\Omega}$.  In this expression
$\delta^{(k,k')}_\varphi T$ denotes the perturbations
(\ref{defperturbationk}) in the gauge $\varphi$.

\subsection{Gauge invariance}\label{gaugeinvariance}

Given a tensor field $T$, if 
$T^\varphi_{\lambda,\Omega}=T^\psi_{\lambda,\Omega}$, 
for any pair of gauges $\varphi$ and $\psi$, we say that $T$ is 
{\em totally gauge invariant}.  In general this is a very strong condition, 
because then (\ref{defexpX}) and the analogous equation in the gauge $\psi$ 
would
imply that $\delta^{(k,k')}_\varphi T= \delta^{(k,k')}_\psi T$, for any two 
gauges
$\varphi$ and $\psi$ and for any order $(k,k')$.  However, in practice one is 
interested
in perturbations up to a fixed order. It is then convenient to weaken the 
definition 
given above, saying that $T$ is {\em gauge invariant up to order $(n,n')$}
iff for any two gauges $\varphi$ and $\psi$ we have
\begin{eqnarray}
\delta^{(k,k')}_\varphi T =\delta^{(k,k')}_\psi T
~~~~~~\forall~~(k,k')~~\hbox{ with}~ k\leq n\,,\,k'\leq n'\,. \nonumber
\end{eqnarray}

From this definition and the definition of perturbation of order $(k,k')$ we
have that a tensor field $T$ is gauge invariant to order $(n,n')$ iff in 
any gauge $\varphi$ we have that ${\cal L}_{\xi}\delta^{(k,k')}_\varphi
T=0$, for any vector field $\xi$ defined on ${\cal M}$ and for any
$(k,k')<(n,n')$.  As a consequence, $T$ is gauge invariant to order $(n,n')$ iff
$T_0$ and all its perturbations of order lower than $(n,n')$ are, in any gauge,
either vanishing or constant scalars, or a combination of Kronecker deltas with
constant coefficients.

\subsection{Gauge transformations}

When a tensor field $T$ is not gauge invariant, it is important to know how its
representation on ${\cal M}_0$ changes under a gauge transformation.   To this
purpose, given two gauges $\varphi$ and $\psi$, it is natural to introduce,
for each value of $(\lambda,\Omega)\in{\rm I\!R}^2$, the diffeomorphism
$\Phi_{\lambda,\Omega}
:{\cal M}_0\rightarrow{\cal M}_0$ defined by
\begin{eqnarray}
\Phi_{\lambda,\Omega}:=\varphi^{-1}_{\lambda,\Omega}\circ\psi_{\lambda,
\Omega}= \varphi_{-\!\lambda,-\!\Omega}\circ\psi_{\lambda,
\Omega}\,, \nonumber
\end{eqnarray}
which represents the gauge transformation from the gauge $\varphi$ to the gauge
$\psi$.  Indeed, the tensor fields $T^\varphi_{\lambda,\Omega}$ and
$T^\psi_{\lambda,\Omega}$, defined on ${\cal M}_0$ by the gauges $\varphi$ and
$\psi$, are connected by the linear map $\Phi_{\lambda,\Omega}^*$, the pull-back
of $\Phi_{\lambda,\Omega}$:
\begin{eqnarray}
T^\psi_{\lambda,\Omega} & = & \left.
\psi^*_{\lambda,\Omega}T\,\right|_{{\cal M}_0}= \left.
\left(\psi^*_{\lambda,\Omega}\varphi^*_{\!-\!\lambda,\!-\!\Omega}
\varphi^*_{\lambda,\Omega}T\right)\right|_{{\cal M}_0} \nonumber \\
& = & \left. \Phi^*_{\lambda,\Omega}\left(\varphi^*_{\lambda,\Omega}T
\right)\right|_{{\cal M}_0} = 
\Phi^*_{\lambda,\Omega}T^\varphi_{\lambda,\Omega}\,.
\label{gaugetrans}
\end{eqnarray}
We must stress that $\Phi:{\cal M}_0\times{\rm I\!R}^2\rightarrow{\cal M}_0$
thus defined, {\em does not} constitute a two--parameter group of 
diffeomorphisms
in ${\cal M}_0$.  In other words, the action of $\Phi$ cannot be generated
by a pair of vector fields.  However, one can show~\cite{BGS} that for
any tensor field $T$, $\Phi_{\lambda,\Omega}^\ast T$ can be expanded up
to total order $n$ (that is, including all the terms or order $(k,k')$ with 
$k+k'
\leq n$) by using $n(n+1)/2$ vector fields.  Up to second order,  the explicit
form of the expansion of equation~(\ref{gaugetrans}) is (details and the 
fourth-order expansion are given in~\cite{BGS})
\begin{eqnarray}
 T^\psi_{\lambda,\Omega} & = & T^\varphi_{\lambda,\Omega}
+\lambda{\cal L}_{\xi_{(1,0)}}T^\varphi_{\lambda,\Omega}+
\Omega{\cal L}_{\xi_{(0,1)}}T^\varphi_{\lambda,\Omega} \nonumber \\
 & + & \frac{\lambda^2}{2}\left\{{\cal L}_{\xi_{(2,0)}}
+{\cal L}^2_{\xi_{(1,0)}}\right\}T^\varphi_{\lambda,\Omega}
+\frac{\Omega^2}{2}\left\{{\cal L}_{\xi_{(0,2)}}+{\cal 
L}^2_{\xi_{(0,1)}}\right\}
T^\varphi_{\lambda,\Omega} \nonumber \\
 & + & \lambda\Omega\left\{{\cal L}_{\xi_{(1,1)}}
+\epsilon_0{\cal L}_{\xi_{(1,0)}}{\cal L}_{\xi_{(0,1)}}+
\epsilon_1{\cal L}_{\xi_{(0,1)}}{\cal L}_{\xi_{(1,0)}}\right\}
T^\varphi_{\lambda,\Omega}  +  O^3(\lambda,\Omega) \,, \nonumber
\end{eqnarray}
where the $\xi_{(p,q)}$ are the vector fields generating the gauge 
transformation
$\Phi_{\lambda,\Omega}$ and $(\epsilon_0,\epsilon_1)$ are two real constants
such that $\epsilon_0+\epsilon_1=1$.  They represent the freedom we have in
the reconstruction of the gauge transformation.

With all these ingredients, we can now relate the perturbations in the two 
gauges
$\varphi$ and $\psi$. To second total order, these relations are given by
\begin{eqnarray}
\delta^{(1,0)}_\psi T-\delta^{(1,0)}_\varphi T = {\cal L}_{\xi_{(1,0)}}T_0\,,
~~~~
\delta^{(0,1)}_\psi T-\delta^{(0,1)}_\varphi T = {\cal L}_{\xi_{(0,1)}}T_0\,,
\label{gi1}
\end{eqnarray}
\begin{eqnarray}
\delta^{(2,0)}_\psi T-\delta^{(2,0)}_\varphi T = 2{\cal L}_{\xi_{(1,0)}}
\delta^{(1,0)}_\varphi T +\left\{{\cal L}_{\xi_{(2,0)}}
+{\cal L}_{\xi_{(1,0)}}^2\right\}T_0\,, \label{gi20}
\end{eqnarray}
\begin{eqnarray}
\delta^{(1,1)}_\psi T-\delta^{(1,1)}_\varphi T = {\cal L}_{\xi_{(1,0)}}
\delta^{(0,1)}_\varphi T + {\cal L}_{\xi_{(0,1)}}\delta_\varphi^{(1,0)}T  
\nonumber \\
 ~~~~~ + \left\{{\cal L}_{\xi_{(1,1)}}+\epsilon_0{\cal L}_{\xi_{(1,0)}}
{\cal L}_{\xi_{(0,1)}}+\epsilon_1{\cal L}_{\xi_{(0,1)}}{\cal L}_{\xi_{(1,0)}}
\right\}T_0 \,,  \nonumber
\end{eqnarray}
\begin{eqnarray}
\delta^{(0,2)}_\psi T-\delta^{(0,2)}_\varphi T = 2{\cal L}_{\xi_{(0,1)}}
\delta^{(0,1)}_\varphi T+\left\{{\cal L}_{\xi_{(0,2)}}
+{\cal L}_{\xi_{(0,1)}}^2\right\}T_0\,, \nonumber
\end{eqnarray}
This result is, of course, consistent with the characterization
of gauge invariance above, in the subsection~\ref{gaugeinvariance}.
Equation (\ref{gi1}) implies that $T$ is gauge invariant to the order
$(1,0)$ or $(0,1)$ iff ${\cal L}_{\xi}T_0=0$ for any vector field
$\xi$ on ${\cal M}_0$. Equation (\ref{gi20}) implies that $T$
is gauge invariant to the order $(2,0)$ iff ${\cal L}_{\xi}T_0=0$ {\em and}
${\cal L}_{\xi}\delta^{(1,0)}_\varphi T=0$ for any vector field $\xi$ on ${\cal
M}_0$, and so on for all the orders.

Finally, we mention that it is also possible to find the explicit expressions
for the generators $\xi_{(p,q)}$ of the gauge transformation $\Phi$ in terms
of the vector fields $({}^\varphi\eta,{}^\varphi\zeta)$
and $({}^\psi\eta,{}^\psi\zeta)$ associated with the gauge choices $\varphi$
and $\psi$ respectively.   Their expressions up to second total order
is~\cite{BGS}:
\begin{eqnarray}
\xi_{(1,0)} = {}^\psi\eta - {}^\varphi\eta \,, ~~~~
\xi_{(0,1)} = {}^\psi\zeta - {}^\varphi\zeta \,, \nonumber
\end{eqnarray}
\begin{eqnarray}
\xi_{(2,0)} = [{}^\varphi\eta,{}^\psi\eta] \,, ~~~~
\xi_{(1,1)} = \epsilon_0[{}^\varphi\eta,{}^\psi\zeta]+
\epsilon_1[{}^\varphi\zeta,{}^\psi\eta] \,, ~~~~
\xi_{(0,2)} = [{}^\varphi\zeta,{}^\psi\zeta]\,. \nonumber
\end{eqnarray}




\begin{thebibliography}{99}

\setlength{\itemsep}{-0.6 ex}   

\bibitem{BMMS} M. Bruni, S. Matarrese, S. Mollerach \& S. Sonego,
{\it Class. Quantum Grav.\/ } {\bf 14}, 2585 (1997)

\bibitem{closelimit} R.H. Price \& J. Pullin, {\it Phys.\ Rev.\ Lett.\/}
{\bf 72}, 3297 (1994); R.J. Gleiser, C.O. Nicasio, R.H. Price \& J. Pullin,
{\it Phys.\ Rev.\ Lett.\/} {\bf 77}, 4483 (1998); G. Khanna, J. Baker,
R.J. Gleiser, P. Laguna, C.O. Nicasio, H. Nollert, R.H. Price \& J. Pullin,
{\it Phys.\ Rev.\ Lett.\/} {\bf 83} 3581 (1999)

\bibitem{BGS} M. Bruni, L. Gualtieri \& C.F. Sopuerta, submitted to
{\it Class. Quantum Grav.\/ } (gr-qc/0207105)

\bibitem{stewart} J.M. Stewart and M. Walker, {\it Proc.\ R.\ Soc.\
London A\/} {\bf 341}, 49 (1974)

\end{thebibliography}
\end{document}